\documentclass[journal]{IEEEtran}

\usepackage{amsmath,amsfonts}
\usepackage{algorithm}
\usepackage{array}
\usepackage[caption=false,font=footnotesize]{subfig}
\usepackage{textcomp}
\usepackage{stfloats}
\usepackage{url}
\usepackage{verbatim}
\usepackage{graphicx}
\usepackage{multirow}
\usepackage{tikz}
\usetikzlibrary{positioning,arrows.meta,fit,calc}

\usepackage{amssymb}
\usepackage{amsthm}
\usepackage{algpseudocode}
\usepackage{bm}
\usepackage{booktabs}

\ifCLASSOPTIONcompsoc
  \usepackage[nocompress]{cite}
\else
  \usepackage{cite}
\fi

\ifCLASSINFOpdf
\else
\fi

\begin{document}

\title{Semantic Sampling via\\
Learnable Observation Front Ends}

\author{Yuxuan~Liu,
        Guangming~Shi,~\IEEEmembership{Fellow,~IEEE,} 
        Pengfei~He,~\IEEEmembership{Graduate Student Member,~IEEE,} 
        Shuai~Ma,~\IEEEmembership{Member,~IEEE,} 
        and~Xiang~Cheng,~\IEEEmembership{Fellow,~IEEE} 
\thanks{This work was supported in part by the National Science and Technology Major Project - Mobile Information Networks under Grant No.2024ZD1300700, and in part by the Natural Science Foundation (NSF) of China No.62293483. \textit{(Corresponding Author: Guangming Shi.)}}
\thanks{Yuxuan Liu is with the School of Electronic and Computer Engineering, Shenzhen Graduate School, Peking University, Shenzhen 518055, China (e-mail: liuyuxuan@stu.pku.edu.cn).}
\thanks{Guangming Shi is with the Peng Cheng Laboratory, Shenzhen 518066, China, and also with the School of Artificial Intelligence, Xidian University, Xi’an, Shaanxi 710071, China (e-mail: gmshi@xidian.edu.cn).}
\thanks{Pengfei He is with the School of Artificial Intelligence, Xidian University, Xi’an, Shaanxi 710071, China (e-mail: hepengfei@stu.xidian.edu.cn).}
\thanks{Shuai Ma is with the Peng Cheng Laboratory, Shenzhen 518066, China (e-mail: mash01@pcl.ac.cn).}
\thanks{Xiang Cheng is with the State Key Laboratory of Photonics and Communications, School of Electronics, Peking University, Beijing 100871, China (e-mail: xiangcheng@pku.edu.cn).}}

\markboth{Preprint}%
{Liu \MakeLowercase{\textit{et al.}}: Semantic Sampling via Learnable Observation Front Ends}

\maketitle

\begin{abstract}
Sampling determines the form of information available to downstream reconstruction systems. Conventional low-rate sampling forms finite-dimensional observations directly from the raw waveform, with the sampling rule mainly guided by bandwidth, sparsity, or fixed signal-level structures. For acoustic signals such as speech, however, reconstruction-relevant information is often expressed through content-related spectral-temporal structures rather than waveform samples alone. This paper proposes semantic sampling via learnable observation front ends, where finite-dimensional observations are generated from learned signal responses instead of directly subsampled waveform points. The proposed front end consists of a semantic feature filterbank, a constrained semantic observation matrix, and a low-rate readout module. The filterbank maps the input waveform into multiple acoustic response channels, the observation matrix combines these responses into a small number of observation channels, and the readout module produces low-rate finite-dimensional samples. A reconstruction network is then used to recover the signal from the resulting observations. Experiments on low-rate speech reconstruction show that, under the same observation budget, the proposed semantic sampling front end provides more informative observations than fixed low-rate sampling and neural restoration methods based on predetermined low-rate waveforms. The improvements in waveform fidelity, spectral consistency, and perceptual quality show that learnable observation front ends preserve more useful information for acoustic signal reconstruction under the same observation budget.

\end{abstract}

\begin{IEEEkeywords}
Semantic sampling, learnable observation front end, signal observation formation, low-rate observation, speech reconstruction, acoustic signal processing.
\end{IEEEkeywords}

\section{Introduction}

\IEEEPARstart{S}{ampling} determines how a continuous signal is represented by finite observations. Classical sampling theory establishes stable recovery conditions mainly from the perspective of bandwidth and signal-space representation~\cite{nyquist1928certain,shannon1949communication,unser2002sampling}. Generalized sampling and finite-rate-of-innovation theory further show that the sampling process may exploit structured signal models beyond pointwise waveform samples~\cite{papoulis1977generalized,vetterli2002sampling}. Compressive sensing extends this principle by demonstrating that sparse or structured signals can be recovered from far fewer measurements than required by conventional Nyquist-rate sampling~\cite{candes2006robust,donoho2006compressed,candes2006stable,candes2008restricted,baraniuk2010model}. These developments share a common insight: the effectiveness of finite observations depends not only on their number, but also on whether the sampling process preserves the structures that are useful for the intended recovery task.

Most existing sampling formulations characterize such useful structures at the signal level, for example through spectral support, sparse coefficients, dictionaries, random projections, or low-dimensional parametric models. These descriptions are powerful and have led to many successful sampling and reconstruction methods. However, acoustic signals such as speech contain rich content-related structures that are not fully described by raw waveform samples or fixed signal-level bases alone. Local spectral envelopes, harmonic patterns, voicing states, energy trajectories, transient components, and time-frequency textures all influence the recoverability and perceptual quality of speech. When the observation budget is limited, direct low-rate waveform sampling may remove reconstruction-relevant acoustic structures before the recovery model can use them. Recent neural audio super-resolution and restoration methods have shown strong reconstruction ability from predetermined low-rate or degraded inputs~\cite{kuleshov2017audio,lee2021nu,han2022nu,liu2021voicefixer,liu2024audiosr,lu2024towards}, yet the input observation in these systems is usually fixed by a preceding sampling or degradation process. Once this observation has been formed, the reconstruction model can only infer missing details from the information that remains.

This paper focuses on the sampling stage: how to form finite-dimensional observations before reconstruction. Instead of sampling the waveform directly, the proposed method first maps the waveform into a learnable acoustic response space and then reads out low-rate observations from that space. We refer to this process as semantic sampling, because the observations are formed from acoustic responses that capture reconstruction-relevant speech structures.

To realize this idea, we propose a learnable observation front end for semantic sampling. The front end consists of a semantic feature filterbank, a constrained semantic observation matrix, and a low-rate readout module. The filterbank produces multiple acoustic response channels from the input waveform. The observation matrix combines these channels into a small number of observation responses under a magnitude-constrained signed structure. The readout module then converts the observation responses into finite-dimensional samples through temporal integration. The resulting observations are temporal readouts of learned acoustic responses, rather than point samples of the raw waveform. A reconstruction network recovers the waveform from these observations, so the reconstruction quality reflects how much useful information the front end preserves. The filter--mixing--readout design keeps the sampling process structured while allowing the observations to adapt to acoustic data and the reconstruction objective. Learnable acoustic front ends have been widely used for recognition and classification~\cite{wang2017trainable,zeghidour2018learning,ravanelli2018speaker,zeghidour2021leaf}. In contrast, the learned responses here are used to form finite-dimensional samples for signal reconstruction.

We evaluate the proposed semantic sampling method on low-rate speech reconstruction, where reconstruction quality directly reflects the information retained in the formed observations. The experiments compare the proposed front end with fixed low-rate sampling and neural restoration methods that operate on predetermined low-rate waveforms under the same observation budget. Results show improvements in waveform fidelity, spectral consistency, and perceptual quality, indicating that the sampling front end directly affects the information available to the reconstruction model.

The main contributions of this work are summarized as follows:
\begin{itemize}
    \item We propose semantic sampling via learnable observation front ends, shifting the sampling object from raw waveform points to learned signal responses that encode reconstruction-relevant structures.
    \item We design a structured observation front end composed of a learnable acoustic filterbank, a constrained semantic observation matrix, and a low-rate readout module. The proposed front end forms finite-dimensional observations through response analysis, channel mixing, and temporal readout.
    \item We validate the proposed method on low-rate speech reconstruction. Under the same observation budget, the proposed semantic observations lead to improved waveform fidelity, spectral consistency, and perceptual quality compared with fixed low-rate sampling and neural restoration baselines.
\end{itemize}

\section{Related Work}

\subsection{Sampling Theory and Learnable Acquisition}

Classical sampling theory studies the recovery of continuous signals from finite observations. The Nyquist--Shannon sampling framework establishes the fundamental relationship between bandwidth and sampling rate for bandlimited signals~\cite{nyquist1928certain,shannon1949communication}. Later developments extended this viewpoint in several directions. Landau studied density conditions for sampling and interpolation~\cite{landau1967necessary}, Papoulis introduced generalized sampling expansion through multiple linear measurements~\cite{papoulis1977generalized}, and Unser provided a systematic account of sampling theory after Shannon, emphasizing stability, generalized sampling, and practical reconstruction issues~\cite{unser2002sampling}. Finite-rate-of-innovation sampling further showed that certain continuous-time signals can be sampled according to their innovation rate rather than their highest frequency~\cite{vetterli2002sampling}. These works indicate that the sampling requirement is determined not only by signal bandwidth, but also by the structural model used to describe the signal.

Compressive sensing provides another important line of work by showing that sparse or structured signals can be recovered from a small number of linear measurements~\cite{candes2006robust,donoho2006compressed,candes2006near,candes2006stable}. The restricted isometry property and related tools offer theoretical guarantees for stable recovery from incomplete and inaccurate observations~\cite{candes2008restricted}. Model-based compressive sensing further incorporates structured sparsity, such as block and tree models, to reduce the number of required measurements~\cite{baraniuk2010model}. Sub-Nyquist sampling of sparse wideband analog signals also connects compressive sampling theory with implementable analog acquisition systems~\cite{mishali2010theory}. These methods establish signal structure as a central factor in sampling design. The structures exploited in this line of work are commonly expressed through spectral support, sparsity, parametric degrees of freedom, or other signal-level descriptions.

Learning-based acquisition methods extend this principle by optimizing the measurement process together with the reconstruction system. ReconNet introduced a convolutional reconstruction network for compressively sensed measurements~\cite{kulkarni2016reconnet}, while DeepCodec jointly optimized sensing and recovery using deep neural networks~\cite{mousavi2017deepcodec}. Other studies learned compressed sensing measurement matrices through gradient unrolling~\cite{wu2019learning}, used generative models as priors for compressed sensing~\cite{bora2017compressed}, or developed scalable and content-aware deep compressed sensing frameworks~\cite{wu2019deep,zhang2021scalable,chen2022content}. In medical imaging, learning-based compressive MRI and learned under-sampling pattern optimization show that sampling patterns can be adapted to data distributions and reconstruction objectives~\cite{gozcu2018learning,bahadir2019learning}. These studies show that the observation process can be optimized jointly with reconstruction. For acoustic sampling, this idea can be realized through a structured front end in which learnable filtering, constrained channel mixing, and low-rate temporal readout jointly determine the finite-dimensional observations.

\subsection{Acoustic Front Ends and Learned Audio Representations}

Acoustic front ends transform raw waveforms into representations that expose useful structures for downstream processing. Conventional front ends such as short-time Fourier analysis and filterbank features rely on predefined time-frequency structures. More recent work has shown that acoustic front ends can be learned directly from data. Trainable front ends have been used to improve keyword spotting under far-field and noisy conditions~\cite{wang2017trainable}. Learning filterbanks from raw speech has been shown to be effective for phone recognition~\cite{zeghidour2018learning}. SincNet constrains the first convolutional layer using parameterized sinc filters, providing an interpretable frequency-selective front end for speaker recognition~\cite{ravanelli2018speaker}. LEAF further unifies learnable filtering, pooling, compression, and normalization into an adaptive audio front end for classification tasks~\cite{zeghidour2021leaf}.

Learned audio representations also provide evidence that acoustic signals can be described through compact latent or discrete structures. Variational autoencoders learn continuous latent variables for generative modeling~\cite{kingma2013auto}, VQ-VAE introduces discrete codebook representations~\cite{van2017neural}, and contrastive predictive coding learns compact representations through prediction-based objectives~\cite{oord2018representation}. Neural audio compression systems, such as high-fidelity neural audio codecs, further demonstrate that audio can be represented and reconstructed through learned intermediate representations~\cite{defossez2022high}. These works show that learned response spaces can organize acoustic information more flexibly than fixed waveform samples. In this work, the learned responses are instead read out as finite-dimensional observations for waveform reconstruction, rather than used as recognition features.

\begin{table*}[t]
    \centering
    \caption{Positioning of the proposed method with respect to related work.}
    \label{tab:related_work_positioning}
    \renewcommand{\arraystretch}{1.2}
    \begin{tabular}{l l c c c l l}
    \hline
    \textbf{Line of work} &
    \textbf{Representative works} &
    \textbf{Finite obs.} &
    \textbf{Acoustic resp.} &
    \textbf{Learnable} &
    \textbf{Prior} &
    \textbf{Target} \\
    \hline
    Classical sampling &
    \cite{nyquist1928certain,shannon1949communication,unser2002sampling} &
    -- &
    -- &
    -- &
    Bandwidth prior &
    Signal reconstruction \\
    
    Compressive sensing &
    \cite{candes2006robust,donoho2006compressed,candes2008restricted,baraniuk2010model} &
    \checkmark &
    -- &
    -- &
    Sparsity prior &
    Signal reconstruction \\
    
    Learned compressed acquisition &
    \cite{kulkarni2016reconnet,mousavi2017deepcodec,wu2019learning,zhang2021scalable,chen2022content} &
    \checkmark &
    -- &
    \checkmark &
    Data prior &
    Signal reconstruction \\
    
    Learnable acoustic front ends &
    \cite{wang2017trainable,zeghidour2018learning,ravanelli2018speaker,zeghidour2021leaf} &
    -- &
    \checkmark &
    \checkmark &
    Acoustic prior &
    Downstream task \\
    
    Neural audio restoration &
    \cite{kuleshov2017audio,lee2021nu,han2022nu,liu2021voicefixer,liu2024audiosr} &
    -- &
    -- &
    \checkmark &
    Restoration prior &
    Signal reconstruction \\
    
    Semantic representations &
    \cite{xie2021deep,qin2021semantic,gunduz2022beyond,shao2024theory} &
    -- &
    -- &
    \checkmark &
    Task prior &
    Downstream task \\
    
    \textbf{Semantic sampling} &
    \textbf{This work} &
    \textbf{\checkmark} &
    \textbf{\checkmark} &
    \textbf{\checkmark} &
    \textbf{Semantic prior} &
    \textbf{Signal reconstruction} \\
    \hline
    \end{tabular}
\end{table*}

\subsection{Semantic and Task-Oriented Representations}

The notion of semantic information has been widely studied in representation learning and communication systems. In semantic communication, the objective is no longer limited to reproducing transmitted symbols or waveforms, but is often related to recovering task-relevant or meaning-related information at the receiver. Deep learning based semantic communication systems, such as DeepSC, learn end-to-end semantic encoding and transmission mechanisms~\cite{xie2021deep}. Surveys and theoretical studies have discussed the principles, challenges, and information-theoretic aspects of semantic and task-oriented communication~\cite{qin2021semantic,gunduz2022beyond,shao2024theory}. Related works have also examined explainable semantic features, rate-distortion-perception tradeoffs, classification-oriented communication, and generative task-oriented semantic transmission~\cite{ma2023task,chai2023rate,kutay2024classification,fu2025generative}.

These studies show that the useful information in a signal is not always identical to the full waveform or symbol sequence. For acoustic sampling, this means that finite observations can focus on speech structures that are important for reconstruction. The proposed front end follows this idea by reading out low-rate observations from learned acoustic responses instead of directly sampling the waveform.

\subsection{Neural Audio Reconstruction from Fixed Observations}

Neural networks have substantially advanced audio super-resolution, bandwidth extension, and speech restoration. AudioUNet showed that convolutional encoder--decoder networks can recover high-resolution audio from low-resolution inputs~\cite{kuleshov2017audio}. Generative adversarial models and neural vocoders have improved waveform generation and bandwidth extension quality~\cite{kim2019bandwidth,yamamoto2020parallel,kong2020hifi}. Diffusion models have also been applied to audio synthesis and neural audio upsampling, leading to methods such as DiffWave, NU-Wave, and NU-Wave 2~\cite{kong2020diffwave,lee2021nu,han2022nu}. VoiceFixer addresses general speech restoration using neural vocoder-based modeling~\cite{liu2021voicefixer}, while AudioSR extends audio super-resolution to a broader range of audio types at scale~\cite{liu2024audiosr}. Recent speech bandwidth extension work has further considered amplitude and phase modeling to improve reconstruction quality and efficiency~\cite{lu2024towards}.

These methods show the strength of neural reconstruction when a low-rate or degraded input has already been given. In the proposed semantic sampling setting, the input to the reconstruction network is not fixed in advance. It is formed by a learnable front end through acoustic response analysis, constrained channel mixing, and low-rate readout under the same observation budget.

Table~\ref{tab:related_work_positioning} compares the proposed method with related lines of work in terms of finite observation formation, acoustic response analysis, learnability, prior structure, and target task. Unlike learnable acoustic front ends for recognition or neural restoration methods for fixed low-rate inputs, the proposed method learns a front end that forms finite-dimensional observations for signal reconstruction.

\section{Methodology}
\label{sec:methodology}

\subsection{Overview of Semantic Sampling}
\label{subsec:method_overview}

\begin{figure*}[t]
    \centering
    \includegraphics[width=\textwidth]{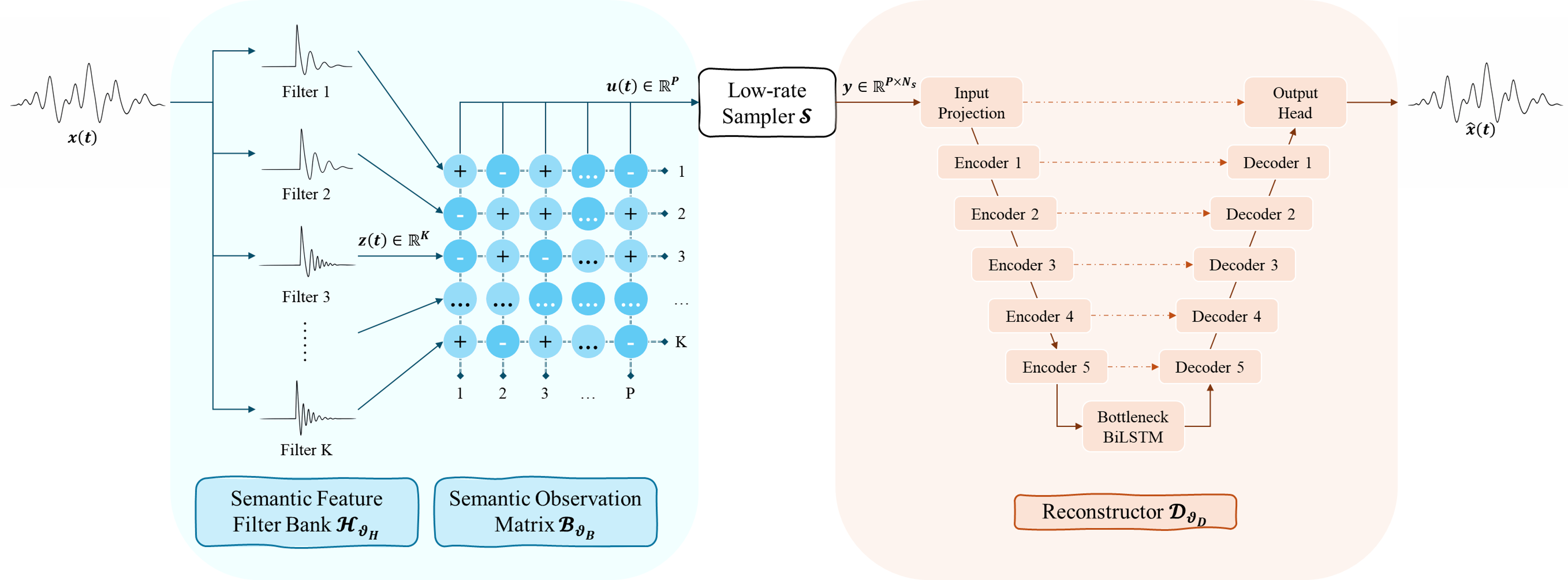}
    \caption{Overview of the proposed semantic sampling framework. The input waveform $x(t)$ is first mapped to multi-channel acoustic responses $z(t)$ by a semantic feature filterbank $\mathcal H_{\vartheta_H}$. A constrained semantic observation matrix $\mathcal B_{\vartheta_B}$ then combines these responses into observation channels $u(t)$. The low-rate readout module $\mathcal S$ converts the observation responses into finite-dimensional samples $y$. A reconstruction network $\mathcal D_{\vartheta_D}$ maps $y$ back to the waveform domain through input projection, an encoder, a BiLSTM bottleneck, a decoder, and an output head.}
    \label{fig:semantic_sampling_framework}
\end{figure*}

Let $x(t)$ denote an input acoustic signal. The goal of semantic sampling is to form a finite-dimensional observation $y$ that preserves acoustic structures useful for reconstructing $x(t)$ under a limited observation budget. The proposed system consists of a learnable sampling front end and a waveform reconstruction network, as illustrated in Fig.~\ref{fig:semantic_sampling_framework}. The overall mapping is written as
\begin{equation}
    y=\mathcal A_{\vartheta_A}(x),
    \qquad
    \hat{x}=\mathcal D_{\vartheta_D}(y),
    \label{eq:method_overall}
\end{equation}
where $\mathcal A_{\vartheta_A}$ is the semantic sampling front end, $\mathcal D_{\vartheta_D}$ is the reconstruction network, and $\vartheta_A$ and $\vartheta_D$ denote their learnable parameters.

The front end is decomposed as
\begin{equation}
    \mathcal A_{\vartheta_A}
    =
    \mathcal S
    \circ
    \mathcal B_{\vartheta_B}
    \circ
    \mathcal H_{\vartheta_H},
    \label{eq:frontend_decomposition}
\end{equation}
where $\mathcal H_{\vartheta_H}$ is a learnable semantic feature filterbank, $\mathcal B_{\vartheta_B}$ is a constrained semantic observation matrix, and $\mathcal S$ is a low-rate readout module. The corresponding signal flow is
\begin{equation}
    x(t)
    \xrightarrow{\mathcal H_{\vartheta_H}}
    z(t)
    \xrightarrow{\mathcal B_{\vartheta_B}}
    u(t)
    \xrightarrow{\mathcal S}
    y
    \xrightarrow{\mathcal D_{\vartheta_D}}
    \hat{x}(t).
    \label{eq:method_pipeline}
\end{equation}

In conventional low-rate waveform sampling, finite observations are taken directly from the waveform, such as point samples or fixed windowed averages. In the proposed framework, the readout is applied after learnable acoustic filtering and constrained channel mixing. Thus, the samples in $y$ are low-rate readouts of learned acoustic responses. This is the sampling mechanism used in this work: the front end first extracts acoustic responses and then forms finite observations from them before reconstruction.

\subsection{Learnable Observation Front End}
\label{subsec:acoustic_observation_formation}

The semantic sampling front end forms observations in three stages: acoustic response analysis, constrained response mixing, and low-rate temporal readout. These stages determine the finite-dimensional observation $y$ received by the reconstruction network.

\subsubsection{Semantic Feature Filterbank}
\label{subsubsec:semantic_filterbank}

The semantic feature filterbank contains $K$ learnable filters. The $k$-th filter is parameterized as
\begin{equation}
    \begin{aligned}
    h_k(t)
    &=
    g_k
    \exp(-2\pi b_k t)
    \cos(2\pi f_k t+\phi_k)
    \mathbf 1_{t\ge 0}, \\
    &\hspace{15em} k=1,\ldots,K .
    \end{aligned}
    \label{eq:semantic_filter}
\end{equation}
where $f_k$ is the center frequency, $b_k$ is the damping parameter, $\phi_k$ is the phase, $g_k$ is the gain, and $\mathbf 1_{t\ge0}$ enforces causality. The response of the input signal to this filter is
\begin{equation}
    z_k(t)=(x*h_k)(t),
    \qquad
    k=1,\ldots,K.
    \label{eq:filter_response}
\end{equation}
All response channels are stacked as
\begin{equation}
    z(t)
    =
    \begin{bmatrix}
    z_1(t)\\
    \vdots\\
    z_K(t)
    \end{bmatrix}
    \in\mathbb R^K.
    \label{eq:filter_response_stack}
\end{equation}

The parameters in \eqref{eq:semantic_filter} allow the filterbank to adapt its frequency selectivity, temporal decay, phase, and gain to the acoustic data and reconstruction objective. In implementation, the filters are represented with finite length and are combined with window truncation, mean correction, and energy normalization to maintain stable responses. The filterbank transforms the waveform into a multi-channel acoustic response space, from which the following stages form finite observations.

\subsubsection{Constrained Semantic Observation Matrix}
\label{subsubsec:semantic_observation_matrix}

The filterbank produces $K$ response channels. To form a smaller number of observation channels, these responses are linearly combined by a semantic observation matrix. Let $P$ denote the number of observation channels. The mixed response is defined as
\begin{equation}
    u(t)=Bz(t),
    \qquad
    B\in\mathbb R^{P\times K},
    \label{eq:semantic_mixing}
\end{equation}
or equivalently,
\begin{equation}
    u_p(t)
    =
    \sum_{k=1}^{K}B_{p,k}z_k(t),
    \qquad
    p=1,\ldots,P.
    \label{eq:semantic_mixing_channel}
\end{equation}
Here, $u(t)\in\mathbb R^P$ denotes the observation response after channel mixing.

The observation matrix is constrained to have fixed-magnitude signed entries:
\begin{equation}
    B_{p,k}
    =
    \frac{1}{\sqrt K}
    \operatorname{sgn}(\widetilde B_{p,k}),
    \label{eq:sign_mixing_matrix}
\end{equation}
where $\widetilde B_{p,k}$ is a learnable latent parameter and $B_{p,k}$ is the entry used in the forward observation process. This form keeps the scale of each observation channel stable and restricts the mixing operation to additive and subtractive combinations of filter responses. Different rows of $B$ produce different signed combinations, allowing the $P$ observation channels to collect complementary information from the $K$ acoustic responses.

\subsubsection{Low-Rate Readout}
\label{subsubsec:low_rate_readout}

After channel mixing, the front end obtains $P$ continuous observation responses $u_p(t)$. The low-rate readout module converts each response into a sequence of finite samples. Let the $n$-th readout window be
\begin{equation}
    I_n=[nT_s,(n+1)T_s],
    \qquad
    n=0,\ldots,N_s-1,
    \label{eq:readout_window}
\end{equation}
where $T_s$ is the readout period and $N_s$ is the number of readout windows for each observation channel. The sample from channel $p$ and window $I_n$ is
\begin{equation}
    \begin{aligned}
    y_{p,n}
    &=
    \frac{1}{T_s}
    \int_{I_n} u_p(t)\,dt, \\
    &\hspace{4em}
    p=1,\ldots,P,\quad n=0,\ldots,N_s-1 .
    \end{aligned}
    \label{eq:low_rate_sampling}
\end{equation}
The complete finite-dimensional observation is
\begin{equation}
    y\in\mathbb R^{P\times N_s},
    \label{eq:observation_tensor}
\end{equation}
where $P$ is the number of observation channels and $N_s$ is the number of readout samples per channel. For a signal segment of duration $\tau$, the observation rate is defined as
\begin{equation}
    f_{\mathrm{obs}}
    =
    \frac{P N_s}{\tau}.
    \label{eq:observation_rate}
\end{equation}
This rate counts the total number of scalar observation values produced per second by the semantic sampling front end. For comparison with waveform sampling at an original sampling rate $f_0$, we also use the normalized observation rate
\begin{equation}
    r_{\mathrm{obs}}
    =
    \frac{f_{\mathrm{obs}}}{f_0}.
    \label{eq:normalized_observation_rate}
\end{equation}

The readout operation determines the observation rate while preserving the structured form of the front end. Since \eqref{eq:low_rate_sampling} is applied to $u_p(t)$ rather than directly to $x(t)$, each observation value contains the effect of acoustic filtering, response mixing, and temporal aggregation.

Combining \eqref{eq:filter_response}, \eqref{eq:semantic_mixing_channel}, and \eqref{eq:low_rate_sampling}, each observation value can be written as
\begin{equation}
    y_{p,n}
    =
    \frac{1}{T_s}
    \int_{I_n}
    \sum_{k=1}^{K}
    B_{p,k}(h_k*x)(t)\,dt.
    \label{eq:combined_observation}
\end{equation}
This expression shows that an observation value is obtained by first analyzing the waveform with learnable acoustic filters, then combining the resulting responses, and finally reading out the combined response over a low-rate temporal window. Equivalently, \eqref{eq:combined_observation} can be written as
\begin{equation}
    y_{p,n}
    =
    \langle x,\psi_{p,n}\rangle,
    \label{eq:structured_observation_template}
\end{equation}
where $\psi_{p,n}$ is induced by the filter responses, the signed mixing coefficients, and the readout window. Thus, the finite observation $y$ is formed by structured acoustic readouts rather than direct waveform samples. During training, the filterbank and observation matrix learn which acoustic responses and channel combinations are most useful for reconstruction.

\subsection{Reconstruction from Semantic Observations}
\label{subsec:reconstruction_network}

The reconstruction network maps the finite observations back to the waveform domain:
\begin{equation}
    \hat{x}=\mathcal D_{\vartheta_D}(y).
    \label{eq:decoder_mapping}
\end{equation}
The input $y\in\mathbb R^{P\times N_s}$ has fewer time samples than the target waveform and contains multiple observation channels. The network first projects the observation channels into an internal feature space and adjusts the temporal resolution for waveform reconstruction.

The reconstruction network follows a time-domain encoder--decoder structure. The input projection transforms the observation channels into feature channels suitable for convolutional processing. The encoder uses one-dimensional convolutional layers to aggregate local temporal patterns and increase the feature dimension. The bottleneck BiLSTM captures long-range temporal dependencies across the low-rate readout sequence~\cite{hochreiter1997long,schuster1997bidirectional}. The decoder restores the temporal resolution and combines multi-scale features through skip connections. The output head generates the reconstructed waveform.

The reconstruction network takes the finite observations produced by the sampling front end as its input. The quality of the reconstructed waveform depends on both the acoustic information contained in $y$ and the ability of $\mathcal D_{\vartheta_D}$ to map these observations back to the waveform domain.

\subsection{Training Objective}
\label{subsec:training_objective}

Given training samples $\{x^{(q)}\}_{q=1}^{N_{\mathrm{tr}}}$, the reconstructed output is
\begin{equation}
    \hat{x}^{(q)}
    =
    \mathcal D_{\vartheta_D}
    \left(
    \mathcal A_{\vartheta_A}(x^{(q)})
    \right).
    \label{eq:training_forward}
\end{equation}
The front end and the reconstruction network are optimized jointly by
\begin{equation}
    \min_{\vartheta_A,\vartheta_D}
    \frac{1}{N_{\mathrm{tr}}}
    \sum_{q=1}^{N_{\mathrm{tr}}}
    \mathcal L_{\mathrm{rec}}
    \left(
    \hat{x}^{(q)},x^{(q)}
    \right)
    +
    \mathcal R_{\mathrm{front}}.
    \label{eq:training_objective}
\end{equation}
Here, $\mathcal L_{\mathrm{rec}}$ measures reconstruction consistency, and $\mathcal R_{\mathrm{front}}$ regularizes the sampling front end.

The reconstruction loss combines time-domain, frequency-domain, and energy-domain constraints:
\begin{equation}
    \mathcal L_{\mathrm{rec}}
    =
    \lambda_t \mathcal L_{\mathrm{time}}
    +
    \lambda_f \mathcal L_{\mathrm{freq}}
    +
    \lambda_e \mathcal L_{\mathrm{energy}}.
    \label{eq:reconstruction_loss}
\end{equation}
The time-domain term $\mathcal L_{\mathrm{time}}$ penalizes waveform differences. The frequency-domain term $\mathcal L_{\mathrm{freq}}$ contains multi-resolution STFT magnitude consistency and complex-spectrum consistency, encouraging the reconstructed waveform to preserve spectral envelopes, harmonic structures, and local time-frequency details. The energy term $\mathcal L_{\mathrm{energy}}$ constrains the overall energy relation between the reconstruction and the reference signal.

The front-end regularization is
\begin{equation}
    \mathcal R_{\mathrm{front}}
    =
    \beta_h\mathcal R_h
    +
    \beta_b\mathcal R_b
    +
    \beta_u\mathcal R_u.
    \label{eq:frontend_regularization}
\end{equation}
The term $\mathcal R_h$ regularizes the filterbank by controlling filter energy, smoothness, and channel diversity. The term $\mathcal R_b$ regularizes the observation matrix to encourage diverse observation channels and stable mixing patterns. The term $\mathcal R_u$ regularizes the observation responses before readout, reducing redundancy among observation channels. The complete training objective is
\begin{equation}
    \mathcal L
    =
    \mathcal L_{\mathrm{rec}}
    +
    \mathcal R_{\mathrm{front}}.
    \label{eq:total_training_loss}
\end{equation}

Through this objective, the acoustic filters, signed observation matrix, low-rate readout, and reconstruction network are optimized as a single observation-formation and reconstruction system. The learned front end determines how finite-dimensional observations are formed, while the reconstruction loss adjusts the acoustic responses and their combinations according to waveform, spectral, and energy consistency.

\section{Experiments}
\label{sec:experiments}

\subsection{Experimental Setup}
\label{subsec:experimental_setup}

\subsubsection{Datasets}
The main experiments are conducted on the LibriSpeech corpus~\cite{panayotov2015librispeech}. Following the official LibriSpeech split, the training set contains $100.6$ hours of speech from $251$ speakers with $28{,}539$ utterances, the validation set contains $2{,}703$ utterances, and the test set contains $2{,}620$ utterances. All signals are processed as 16-kHz single-channel waveforms.

AISHELL-1 is used for cross-dataset evaluation~\cite{bu2017aishell}. It contains Mandarin speech recorded from a different speaker population and linguistic distribution from LibriSpeech. The official training, validation, and test splits of AISHELL-1 are used in the cross-dataset experiments. All AISHELL-1 signals are also processed as 16-kHz single-channel waveforms.

During training, segments are randomly cropped from each utterance. Each input segment therefore contains $16000$ waveform samples at the original sampling rate. During validation and testing, fixed center cropping is used, with zero padding applied when necessary. Random gain perturbation within $3$ dB and additive Gaussian noise with standard deviation $5\times10^{-4}$ are applied during training.

\subsubsection{Observation rates}
All methods are compared under the same observation rate. In the proposed semantic sampling front end, the observation rate $f_{\mathrm{obs}}$ counts the total number of scalar observation values produced per second across all observation channels. For waveform-based baselines, $f_{\mathrm{obs}}$ corresponds to the sampling rate of the low-rate waveform input. This setting ensures that all methods receive the same number of scalar input values per second before reconstruction.

We evaluate three observation rates: $4$ kHz, $2$ kHz, and $1$ kHz. Since the reference speech is sampled at $f_0=16$ kHz, these rates correspond to normalized observation rates of $0.25$, $0.125$, and $0.0625$, respectively. The lowest observation rate represents a highly compressed setting, where the finite observations must preserve reconstruction-relevant acoustic structures under a severe input constraint.

\subsubsection{Baselines}
The proposed method is compared with fixed waveform sampling and representative neural audio super-resolution methods. Uniform-Sinc first uniformly downsamples the waveform to the target observation rate and then reconstructs the waveform by sinc interpolation, providing a classical fixed-sampling reference~\cite{shannon1949communication,unser2002sampling}. AudioUNet uses a convolutional encoder--decoder network for audio super-resolution from low-rate waveform inputs~\cite{kuleshov2017audio}. NU-Wave 2 applies a diffusion-based neural audio upsampling model that supports multiple input sampling rates~\cite{han2022nu}. AudioSR is a large-scale diffusion-based audio super-resolution model designed for general audio bandwidth extension~\cite{liu2024audiosr}.

For AudioUNet, NU-Wave 2, and AudioSR, the publicly released pretrained models are used without retraining on the LibriSpeech training set. For these baselines, the original speech is first downsampled to the corresponding observation rate and then reconstructed to 16 kHz by the baseline model. For the proposed method, the semantic sampling front end forms finite-dimensional observations at the same observation rate, and the reconstruction network maps these observations back to the waveform domain. The comparison therefore keeps the scalar observation budget fixed while varying the way in which the low-rate input observations are formed.

In the cross-dataset evaluation on AISHELL-1, all systems are evaluated without retraining, fine-tuning, or parameter adaptation on AISHELL-1. The proposed model uses the parameters learned from LibriSpeech, while the baseline systems use their publicly released pretrained parameters. This protocol evaluates reconstruction from low-rate observations under a different speech corpus and linguistic distribution.

\subsubsection{Evaluation metrics}
The reconstructed signals are evaluated from waveform fidelity, spectral consistency, and perceptual quality. 

Waveform fidelity is measured by scale-invariant signal-to-distortion ratio (SI-SDR)~\cite{le2019sdr} and signal-to-noise ratio (SNR). SI-SDR evaluates the consistency between the reconstructed and reference waveforms after removing scale differences, while SNR measures the reconstruction error relative to the reference signal energy.

Spectral consistency is measured by multi-resolution short-time Fourier transform distance (MR-STFT) and log-spectral distance (LSD). MR-STFT compares the reconstructed and reference spectra under multiple time-frequency resolutions, while LSD measures the average discrepancy in log-magnitude spectra and is sensitive to spectral envelopes, harmonic structures, and band-energy distributions~\cite{yamamoto2020parallel,kong2020hifi,gray2003distance,quackenbush1988objective}.

Perceptual quality is evaluated by short-time objective intelligibility (STOI)~\cite{taal2011algorithm} and perceptual evaluation of speech quality (PESQ)~\cite{recommendation2001perceptual}. Higher SI-SDR, SNR, STOI, and PESQ indicate better reconstruction quality, while lower MR-STFT and LSD indicate better spectral consistency.

\subsubsection{Implementation details}
The proposed semantic sampling front end and reconstruction network are trained end to end. AdamW is used as the optimizer~\cite{loshchilov2017decoupled}. The model is trained for $80$ epochs with a batch size of $64$. The learning rate is set to $5\times10^{-5}$ for the semantic feature filterbank, semantic observation matrix, and low-rate readout module, and to $2\times10^{-4}$ for the reconstruction network. A cosine annealing learning-rate schedule is used, with a minimum learning rate of $10^{-6}$~\cite{loshchilov2016sgdr}. The gradient norm is clipped at $3.0$ for stable training.

The training objective follows \eqref{eq:total_training_loss}. The reconstruction loss contains time-domain, frequency-domain, and energy-domain terms. The time-domain term uses an $\ell_1$ waveform loss. The frequency-domain term combines multi-resolution STFT magnitude consistency and complex-spectrum consistency. The energy-domain term constrains the global energy relation between the reconstructed and reference waveforms. The weights of the three reconstruction terms are set to $\lambda_t=1.0$, $\lambda_f=1.0$, and $\lambda_e=0.05$. The front-end regularization terms are used to stabilize the filterbank, the observation matrix, and the observation responses during joint optimization.

\subsection{Quantitative Results}
\label{subsec:quantitative_results}

Table~\ref{tab:main_quantitative_results} compares reconstruction performance at 4 kHz, 2 kHz, and 1 kHz observation rates. All observation rates denote the total number of scalar observations per second. For 16-kHz speech, 4 kHz, 2 kHz, and 1 kHz correspond to normalized observation rates of 0.25, 0.125, and 0.0625, respectively. Higher SI-SDR, SNR, STOI, and PESQ indicate better performance, while lower MR-STFT and LSD indicate better performance.

\begin{table*}[t]
    \centering
    \caption{Quantitative Comparison at Different Observation Rates}
    \label{tab:main_quantitative_results}
    \begin{tabular}{llcccccc}
        \hline
        \multirow{2}{*}{Observation rate}
        & \multirow{2}{*}{Method}
        & \multicolumn{2}{c}{Waveform fidelity}
        & \multicolumn{2}{c}{Spectral consistency}
        & \multicolumn{2}{c}{Perceptual quality} \\
        \cline{3-8}
        & & SI-SDR $\uparrow$ & SNR $\uparrow$
        & MR-STFT $\downarrow$ & LSD $\downarrow$
        & STOI $\uparrow$ & PESQ $\uparrow$ \\
        \hline
        \multirow{5}{*}{4 kHz}
        & Uniform-Sinc & 11.01 & 11.59 & 4.82 & 46.69 & 0.82 & 1.87 \\
        & AudioUNet    & 3.67  & 3.58  & 2.22 & 16.57 & 0.89 & 2.31 \\
        & NU-Wave 2    & 12.60 & 12.95 & 1.54 & 14.92 & 0.88 & 2.04 \\
        & AudioSR      & 9.10  & 9.55  & 1.53 & 14.65 & 0.89 & 1.73 \\
        & Ours         & \textbf{15.83} & \textbf{16.01} & \textbf{1.03} & \textbf{10.71} & \textbf{0.95} & \textbf{2.76} \\
        \hline
        \multirow{5}{*}{2 kHz}
        & Uniform-Sinc & 5.25  & 6.69  & 6.28 & 55.59 & 0.73 & 1.50 \\
        & AudioUNet    & 8.09  & 9.07  & 3.02 & 27.23 & 0.79 & 1.86 \\
        & NU-Wave 2    & 8.04  & 9.01  & 1.89 & 16.57 & 0.77 & 1.79 \\
        & AudioSR      & 7.94  & 8.86  & 4.71 & 42.28 & 0.77 & 2.17 \\
        & Ours         & \textbf{11.07} & \textbf{11.52} & \textbf{1.22} & \textbf{11.44} & \textbf{0.91} & \textbf{2.33} \\
        \hline
        \multirow{5}{*}{1 kHz}
        & Uniform-Sinc & -3.78 & 1.66 & 7.54 & 62.58 & 0.63 & 1.24 \\
        & AudioUNet    & 0.45  & 3.98 & 4.26 & 35.60 & 0.65 & 1.41 \\
        & NU-Wave 2    & 0.63  & 4.14 & 2.33 & 18.08 & 0.63 & 1.43 \\
        & AudioSR      & 0.48  & 4.03 & 6.35 & 52.73 & 0.65 & 1.73 \\
        & Ours         & \textbf{7.13} & \textbf{8.09} & \textbf{1.45} & \textbf{12.47} & \textbf{0.84} & \textbf{1.74} \\
        \hline
    \end{tabular}
\end{table*}

The proposed method achieves the best result on all metrics at all three observation rates. The gains appear in waveform, spectral, and perceptual metrics, showing that the learned acoustic observations retain information useful for different aspects of speech reconstruction.

At the 4 kHz observation rate, NU-Wave 2 and AudioSR provide strong neural restoration baselines, especially in SI-SDR, SNR, and spectral consistency. The proposed method further improves SI-SDR from 12.60 dB to 15.83 dB and SNR from 12.95 dB to 16.01 dB compared with the strongest baseline. It also reduces MR-STFT from 1.53 to 1.03 and LSD from 14.65 to 10.71. These gains show that, even when the observation rate is relatively moderate, forming observations from learned acoustic responses provides a more favorable input for waveform reconstruction than using a fixed low-rate waveform followed by neural restoration.

The advantage remains clear at lower observation rates. At 2 kHz, the strongest baseline reaches 8.09 dB SI-SDR and 9.07 dB SNR, while the proposed method obtains 11.07 dB and 11.52 dB, respectively. The spectral gap is also substantial: the proposed method achieves 1.22 MR-STFT and 11.44 LSD, compared with 1.89 MR-STFT and 16.57 LSD for the strongest spectral baseline. At 1 kHz, where the number of available observation values is severely limited, the proposed method still reaches 7.13 dB SI-SDR and 8.09 dB SNR. The best baseline remains below 1 dB SI-SDR and below 4.2 dB SNR. This large separation suggests that the benefit of acoustic observation formation becomes especially important when direct low-rate waveform observations contain limited recoverable structure.

Across the three observation rates, the proposed method obtains the lowest MR-STFT and LSD values, showing more accurate reconstruction of spectral envelopes, harmonic regions, and band-energy distributions. The STOI values of 0.95, 0.91, and 0.84 show better preservation of intelligibility-related structures as the observation rate decreases. PESQ is also the highest among all methods, although the margin at 1 kHz is small. These results show that the proposed front end improves reconstruction under the same scalar observation budget by changing how the observations are formed, not by increasing the number of input values.

\subsection{Qualitative Results}
\label{subsec:qualitative_results}

\begin{figure*}[t]
    \centering
    \includegraphics[width=\textwidth]{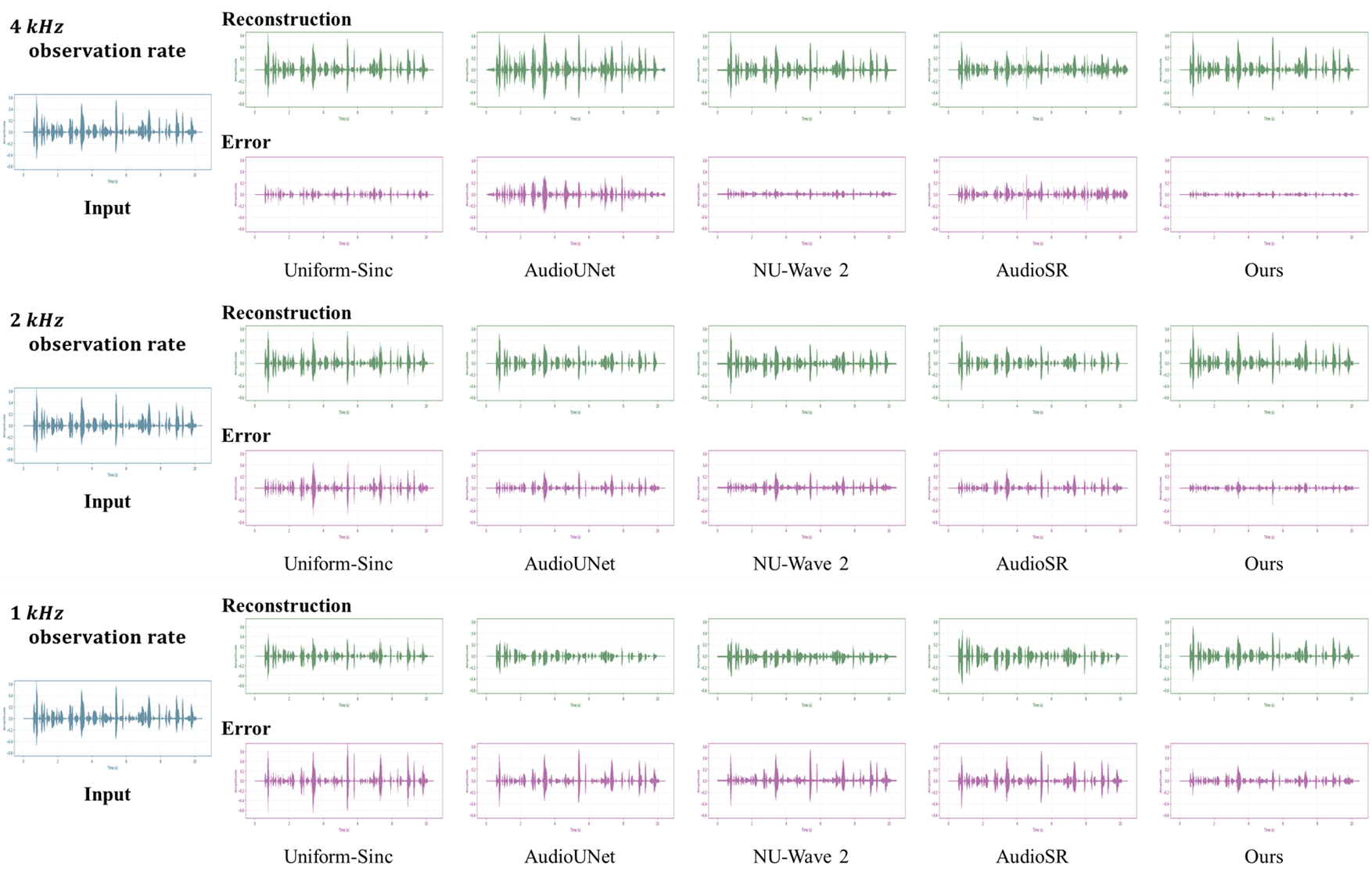}
    \caption{Waveform reconstruction results at different observation rates. Each row corresponds to one observation rate. The left column shows the reference waveform, and the following columns show the reconstructed waveforms and corresponding errors of Uniform-Sinc, AudioUNet, NU-Wave 2, AudioSR, and the proposed method.}
    \label{fig:qual_waveform}
\end{figure*}

Fig.~\ref{fig:qual_waveform} shows waveform reconstruction results and the corresponding errors at different observation rates. The proposed method preserves the main temporal structure of the reference speech across $4$ kHz, $2$ kHz, and $1$ kHz observation rates, including voiced regions, major energy variations, and local waveform peaks. Uniform-Sinc produces increasingly large errors as the observation rate decreases, with visible error spikes around rapidly varying waveform regions. AudioUNet, NU-Wave 2, and AudioSR generate continuous waveforms from low-rate inputs, but their reconstructions show deviations in local amplitude details and fast temporal transitions. In comparison, the proposed method produces smaller error amplitudes and shows a smoother degradation trend as the observation rate decreases, indicating that the semantic sampling front end retains acoustic information useful for recovering the main temporal structure.

\begin{figure*}[t]
    \centering
    \includegraphics[width=\textwidth]{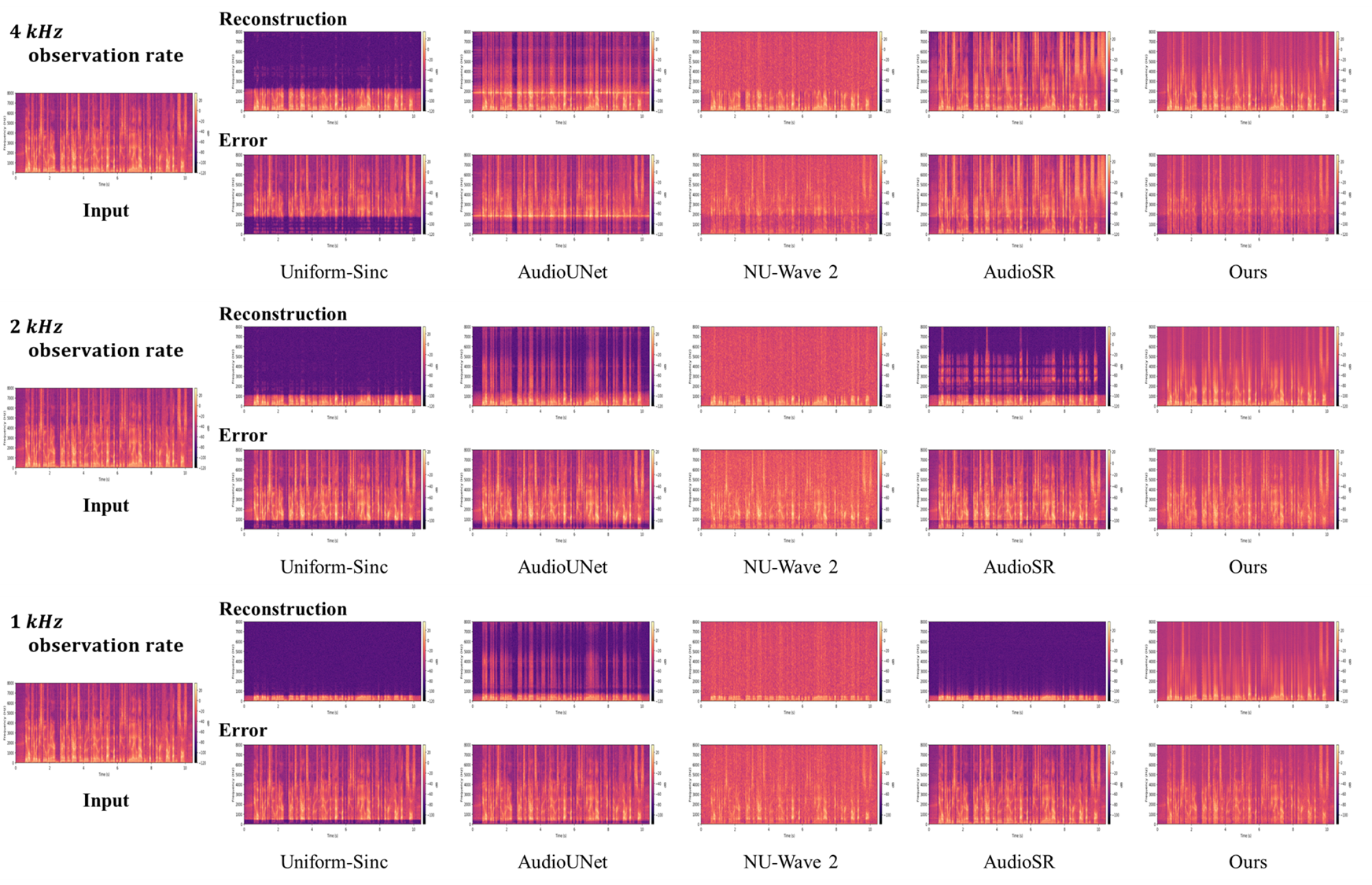}
    \caption{Spectrogram reconstruction results at different observation rates. Each row corresponds to one observation rate. The left column shows the reference spectrogram, and the following columns show the reconstructed spectrograms and corresponding spectral errors of Uniform-Sinc, AudioUNet, NU-Wave 2, AudioSR, and the proposed method.}
    \label{fig:qual_spectrogram}
\end{figure*}

Fig.~\ref{fig:qual_spectrogram} presents the spectrograms of the same speech segment. The proposed method better preserves the main energy bands, harmonic structures, and time-varying spectral patterns. Uniform-Sinc mainly retains low-frequency components at low observation rates, and its spectral error contains substantial residual energy from missing high-frequency and local spectral structures. The neural audio super-resolution baselines recover part of the missing spectral content, but their results may contain blurred spectral patterns, noise-like textures, or generated details that are not well aligned with the reference. The proposed method produces spectrograms that are more consistent with the reference in the main energy distribution, with weaker spectral errors across voiced regions and harmonic bands. This result is consistent with the quantitative improvements in MR-STFT and LSD.

As the observation rate decreases, fixed waveform sampling loses high-frequency structure and local waveform details. Neural restoration methods can synthesize continuous waveforms and partial spectral content, but their reconstructions become less aligned with the reference when the input waveform is severely undersampled. At 1 kHz, the proposed method still preserves the main temporal envelope and dominant spectral distribution. Its errors mainly appear around fast local variations and weaker frequency bands. This result shows the benefit of structured acoustic readouts before reconstruction.

\begin{figure}[t]
    \centering

    \subfloat[Learned filter responses.\label{fig:frontend_probe_filters}]{
        \includegraphics[width=0.95\linewidth]{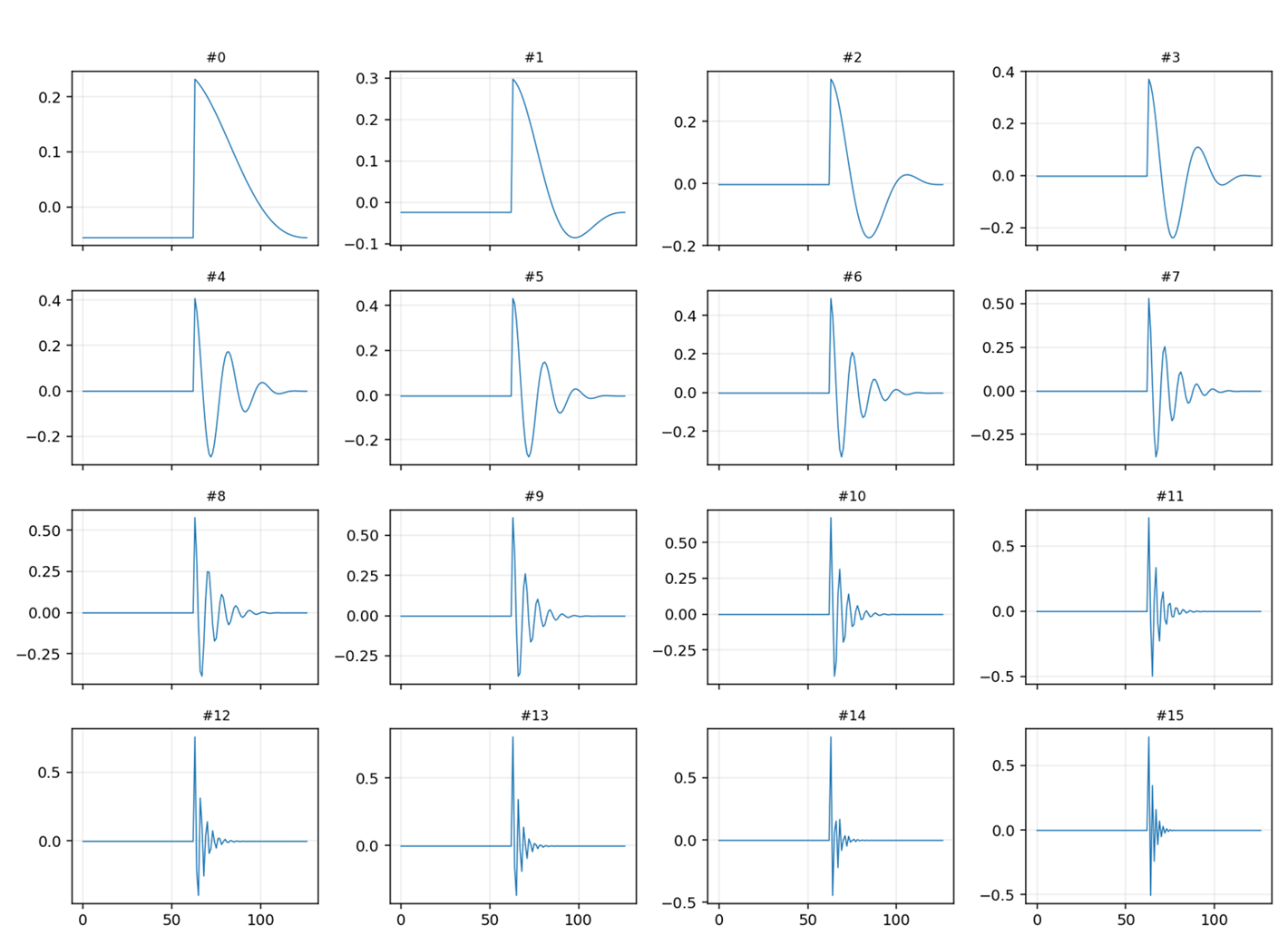}
    }

    \subfloat[Signed semantic observation matrix.\label{fig:frontend_mixing_matrix}]{
        \includegraphics[width=0.95\linewidth]{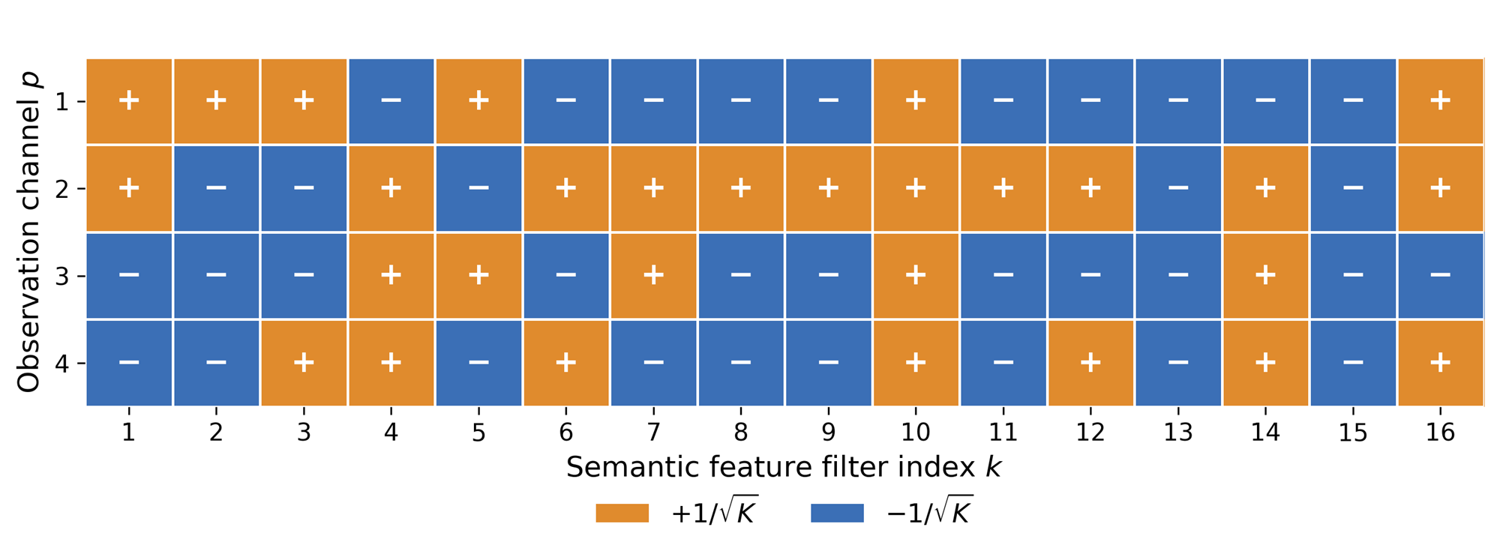}
    }

    \caption{Visualization of the semantic sampling front end at the $2$ kHz observation rate with $K=16$ filter channels and $P=4$ observation channels. Fig.~\ref{fig:frontend_probe_filters} shows the temporal responses of the learned semantic feature filters. Fig.~\ref{fig:frontend_mixing_matrix} shows the signed structure of the semantic observation matrix, where rows correspond to observation channels and columns correspond to filter channels.}
    \label{fig:frontend_visualization}
\end{figure}

Fig.~\ref{fig:frontend_visualization} visualizes the learned front end at the $2$ kHz observation rate. The learned filters exhibit different temporal shapes, frequency selectivity, and decay patterns, forming a multi-channel acoustic response space rather than repeated copies of the same waveform response. The semantic observation matrix has a fixed-magnitude signed structure, where each observation channel combines the filter responses with a different sign pattern. This structure allows the front end to collect complementary acoustic responses into a small number of observation channels before low-rate readout. Together with the waveform and spectrogram results, this visualization shows that the proposed front end forms low-rate observations through acoustic response analysis and constrained channel mixing, which gives the reconstruction network more informative inputs under the same observation rate.

\subsection{Ablation Studies}
\label{subsec:ablation_studies}

Table~\ref{tab:ablation_structure} studies how the front-end configuration affects reconstruction quality. Here, $K$ denotes the number of semantic feature filters, $P$ denotes the number of observation channels, and $L_{\mathrm r}$ denotes the readout interval measured in original waveform samples. Since the original waveform sampling rate is $f_0=16$ kHz, the readout period in \eqref{eq:readout_window} can be written as $T_s=L_{\mathrm r}/f_0$, and the resulting observation rate is
\begin{equation}
    f_{\mathrm{obs}}
    =
    \frac{P}{T_s}
    =
    \frac{P f_0}{L_{\mathrm r}}.
    \label{eq:ablation_observation_rate}
\end{equation}
For each observation-rate block, increasing $P$ is therefore accompanied by a larger $L_{\mathrm r}$ so that the total number of scalar observations per second remains fixed.

\begin{table*}[t]
    \centering
    \caption{Ablation on Front-End Configuration}
    \label{tab:ablation_structure}
    \begin{tabular}{cccc|cc|cc|cc}
        \toprule
        \multirow{2}{*}{Observation rate}
        & \multirow{2}{*}{$K$}
        & \multirow{2}{*}{$P$}
        & \multirow{2}{*}{$L_{\mathrm r}$}
        & \multicolumn{2}{c|}{Waveform fidelity}
        & \multicolumn{2}{c|}{Spectral consistency}
        & \multicolumn{2}{c}{Perceptual quality} \\
        \cmidrule(lr){5-6}\cmidrule(lr){7-8}\cmidrule(lr){9-10}
        & & & &
        SI-SDR $\uparrow$ & SNR $\uparrow$
        & MR-STFT $\downarrow$ & LSD $\downarrow$
        & STOI $\uparrow$ & PESQ $\uparrow$ \\
        \midrule
        \multirow{16}{*}{4 kHz}
        & 16  & 1 & 4  & 14.52 & 14.79 & 1.07 & 10.78 & 0.92 & 2.41 \\
        &     & 2 & 8  & 14.78 & 15.01 & 1.08 & 10.93 & 0.94 & 2.61 \\
        &     & 4 & 16 & 15.32 & 15.54 & 1.08 & 10.98 & 0.94 & 2.69 \\
        &     & 8 & 32 & 11.24 & 11.64 & 1.27 & 11.94 & 0.91 & 2.22 \\
        & 32  & 1 & 4  & 14.52 & 14.78 & 1.07 & 10.84 & 0.93 & 2.49 \\
        &     & 2 & 8  & 14.49 & 14.75 & 1.09 & 10.97 & 0.94 & 2.60 \\
        &     & 4 & 16 & 15.18 & 15.38 & 1.07 & 10.92 & \underline{0.95} & \textbf{2.77} \\
        &     & 8 & 32 & 11.94 & 12.36 & 1.22 & 11.63 & 0.92 & 2.32 \\
        & 64  & 1 & 4  & 15.51 & \underline{15.72} & \textbf{1.03} & \textbf{10.63} & 0.94 & 2.59 \\
        &     & 2 & 8  & \textbf{15.83} & \textbf{16.01} & \underline{1.03} & \underline{10.71} & \textbf{0.95} & 2.76 \\
        &     & 4 & 16 & 13.60 & 13.93 & 1.17 & 11.47 & 0.93 & 2.38 \\
        &     & 8 & 32 & 14.35 & 14.58 & 1.11 & 11.05 & 0.94 & 2.72 \\
        & 128 & 1 & 4  & 14.17 & 14.45 & 1.10 & 11.03 & 0.92 & 2.31 \\
        &     & 2 & 8  & \underline{15.52} & 15.70 & 1.04 & 10.74 & 0.95 & 2.74 \\
        &     & 4 & 16 & 14.97 & 15.17 & 1.10 & 11.14 & 0.94 & 2.62 \\
        &     & 8 & 32 & 14.07 & 14.32 & 1.09 & 10.86 & 0.94 & \underline{2.77} \\
        \midrule
        \multirow{16}{*}{2 kHz}
        & 16  & 1 & 8  & 8.39  & 9.25  & 1.40 & 12.28 & 0.85 & 1.79 \\
        &     & 2 & 16 & 9.81  & 10.43 & 1.30 & 11.91 & 0.88 & 2.00 \\
        &     & 4 & 32 & \textbf{11.07} & \textbf{11.52} & \textbf{1.22} & \textbf{11.44} & \textbf{0.91} & \textbf{2.33} \\
        &     & 8 & 64 & 6.90  & 8.05  & 1.49 & 12.69 & 0.82 & 1.66 \\
        & 32  & 1 & 8  & 8.74  & 9.51  & 1.33 & 11.79 & 0.87 & 1.92 \\
        &     & 2 & 16 & 9.51  & 10.15 & 1.30 & 11.85 & 0.88 & 2.05 \\
        &     & 4 & 32 & 8.54  & 9.35  & 1.40 & 12.39 & 0.85 & 1.80 \\
        &     & 8 & 64 & 7.08  & 8.19  & 1.48 & 12.59 & 0.82 & 1.66 \\
        & 64  & 1 & 8  & 8.62  & 9.42  & 1.34 & 11.87 & 0.86 & 1.86 \\
        &     & 2 & 16 & 9.41  & 10.09 & 1.31 & 11.84 & 0.88 & 2.01 \\
        &     & 4 & 32 & \underline{10.26} & \underline{10.81} & 1.25 & 11.53 & 0.90 & 2.22 \\
        &     & 8 & 64 & 10.08 & 10.63 & \underline{1.24} & \underline{11.44} & \underline{0.91} & \underline{2.28} \\
        & 128 & 1 & 8  & 8.55  & 9.38  & 1.34 & 11.84 & 0.86 & 1.85 \\
        &     & 2 & 16 & 10.01 & 10.62 & 1.25 & 11.45 & 0.90 & 2.18 \\
        &     & 4 & 32 & 8.85  & 9.60  & 1.37 & 12.25 & 0.86 & 1.89 \\
        &     & 8 & 64 & 6.98  & 8.08  & 1.50 & 12.74 & 0.82 & 1.65 \\
        \midrule
        \multirow{16}{*}{1 kHz}
        & 16  & 1 & 16  & 3.52 & 5.49 & 1.72 & 13.76 & 0.77 & 1.35 \\
        &     & 2 & 32  & 4.79 & 6.39 & 1.70 & 13.94 & 0.77 & 1.40 \\
        &     & 4 & 64  & 6.54 & 7.65 & 1.50 & 12.64 & 0.82 & 1.67 \\
        &     & 8 & 128 & 6.56 & 7.66 & 1.49 & 12.55 & 0.82 & 1.67 \\
        & 32  & 1 & 16  & 5.87 & 7.12 & 1.56 & 13.07 & 0.81 & 1.57 \\
        &     & 2 & 32  & 6.54 & 7.64 & 1.50 & 12.67 & 0.83 & 1.67 \\
        &     & 4 & 64  & 6.53 & 7.60 & 1.49 & 12.58 & 0.82 & 1.69 \\
        &     & 8 & 128 & 6.34 & 7.46 & \textbf{1.44} & \textbf{12.07} & 0.83 & 1.72 \\
        & 64  & 1 & 16  & 5.86 & 7.14 & 1.53 & 12.80 & 0.82 & 1.60 \\
        &     & 2 & 32  & \underline{6.89} & \underline{7.90} & 1.47 & 12.54 & \underline{0.83} & 1.71 \\
        &     & 4 & 64  & 0.84 & 3.77 & 1.87 & 14.21 & 0.71 & 1.21 \\
        &     & 8 & 128 & 1.33 & 4.20 & 1.74 & 13.31 & 0.74 & 1.28 \\
        & 128 & 1 & 16  & 6.02 & 7.29 & 1.51 & 12.66 & 0.83 & 1.62 \\
        &     & 2 & 32  & \textbf{7.13} & \textbf{8.09} & \underline{1.45} & \underline{12.47} & \textbf{0.84} & \textbf{1.74} \\
        &     & 4 & 64  & 6.82 & 7.70 & 1.47 & 12.48 & 0.83 & \underline{1.73} \\
        &     & 8 & 128 & 0.06 & 3.39 & 1.83 & 13.60 & 0.72 & 1.21 \\
        \bottomrule
    \end{tabular}
\end{table*}

The ablation results show a tradeoff between acoustic response diversity and temporal readout density. At a fixed observation rate, increasing the number of observation channels $P$ provides more mixed response streams, but it also increases the readout interval $L_{\mathrm r}$. As a result, the additional channel diversity can be offset by a loss of temporal resolution. This effect is visible across the table: configurations with very large $P$ often become less stable, especially when the observation rate is low. For example, at 2 kHz, increasing $P$ from $4$ to $8$ reduces SI-SDR for $K=16$ from $11.07$ dB to $6.90$ dB, and similar drops appear for $K=32$ and $K=128$. At 1 kHz, several $P=8$ configurations show clear degradation in waveform fidelity. These results indicate that increasing the number of observation channels is useful only when each channel can still be read out with sufficient temporal density.

Moderate values of $P$ are more robust than extreme allocations. Configurations with $P=2$ or $P=4$ frequently achieve strong results across waveform, spectral, and perceptual metrics, while $P=1$ can limit the diversity of mixed acoustic responses and $P=8$ often makes the readout too sparse. This pattern is consistent with the role of the semantic observation matrix: multiple observation channels are needed to combine filter responses in complementary ways, but the benefit of additional channels must be balanced against the temporal spacing introduced by the low-rate readout. The observation front end therefore benefits from a balanced allocation between channel diversity and temporal sampling density.

The filterbank size $K$ affects the richness of the acoustic response space, but its effect also depends on the readout allocation. A small filterbank may restrict the range of spectral-temporal responses available for observation formation. For instance, at 1 kHz and $P=2$, increasing $K$ from $16$ to $128$ improves SI-SDR from $4.79$ dB to $7.13$ dB. However, a larger filterbank cannot compensate for an overly sparse readout. When $P$ is large and $L_{\mathrm r}$ becomes too large, performance may degrade even with a large $K$, as shown by the $P=8$ configurations at 1 kHz. Thus, increasing the filterbank capacity is beneficial when the resulting responses can be effectively mixed and read out, but redundant or insufficiently sampled responses do not consistently improve reconstruction.

The ablation study shows that effective semantic sampling requires both a sufficiently expressive acoustic response space and a readout pattern that preserves local temporal structure. The filterbank provides candidate acoustic responses, the observation matrix combines them into complementary observation channels, and the low-rate readout determines how densely these responses are sampled over time. Configurations with balanced channel diversity and readout density produce more informative finite observations, leading to better waveform fidelity, spectral consistency, and perceptual quality. This behavior is consistent with the observation-formation process in Section~\ref{subsec:acoustic_observation_formation}, where each observation value is jointly determined by acoustic filtering, constrained channel mixing, and temporal readout.

\subsection{Cross-Dataset Evaluation}
\label{subsec:cross_dataset_evaluation}

The cross-dataset evaluation is conducted on AISHELL-1~\cite{bu2017aishell}, a Mandarin speech corpus with a different linguistic distribution and speaker population from LibriSpeech. Table~\ref{tab:aishell_cross_dataset_results} reports the results at 4 kHz, 2 kHz, and 1 kHz observation rates. For Uniform-Sinc, AudioUNet, NU-Wave 2, AudioSR, and ``Ours (without fine-tuning)'', no AISHELL-1 training or fine-tuning is performed. The row ``Ours (fine-tuned)'' reports the same proposed model after fine-tuning on the AISHELL-1 training split.

\begin{table*}[t]
    \centering
    \caption{Cross-Dataset Results on AISHELL-1}
    \label{tab:aishell_cross_dataset_results}
    \begin{tabular}{cl|cc|cc|cc}
        \toprule
        \multirow{2}{*}{Observation rate}
        & \multirow{2}{*}{Method}
        & \multicolumn{2}{c|}{Waveform fidelity}
        & \multicolumn{2}{c|}{Spectral consistency}
        & \multicolumn{2}{c}{Perceptual quality} \\
        \cmidrule(lr){3-4}\cmidrule(lr){5-6}\cmidrule(lr){7-8}
        &
        & SI-SDR $\uparrow$ & SNR $\uparrow$
        & MR-STFT $\downarrow$ & LSD $\downarrow$
        & STOI $\uparrow$ & PESQ $\uparrow$ \\
        \midrule
        \multirow{6}{*}{4 kHz}
        & Uniform-Sinc              & 14.06 & 14.28 & 4.61 & 46.28 & 0.83 & 1.93 \\
        & AudioUNet                 & 3.63  & 3.20  & 2.35 & 17.26 & 0.88 & \underline{2.53} \\
        & NU-Wave 2                 & 15.18 & 15.28 & 1.32 & 13.66 & 0.87 & 2.19 \\
        & AudioSR                   & 11.21 & 11.30 & 1.35 & 13.54 & 0.89 & 1.91 \\
        & Ours (without fine-tuning)& \underline{16.61} & \underline{16.67} & \underline{1.12} & \underline{12.12} & \underline{0.93} & 2.47 \\
        \cmidrule(lr){2-8}
        & Ours (fine-tuned)         & \textbf{17.72} & \textbf{17.79} & \textbf{1.03} & \textbf{11.32} & \textbf{0.95} & \textbf{2.73} \\
        \midrule
        \multirow{6}{*}{2 kHz}
        & Uniform-Sinc              & 7.13  & 8.05  & 6.02 & 54.33 & 0.70 & 1.46 \\
        & AudioUNet                 & 9.76  & 10.42 & 2.85 & 26.31 & 0.77 & 2.03 \\
        & NU-Wave 2                 & 9.60  & 10.24 & 1.69 & 15.43 & 0.74 & 1.85 \\
        & AudioSR                   & 9.54  & 10.07 & 4.63 & 42.29 & 0.74 & \textbf{2.30} \\
        & Ours (without fine-tuning)& \underline{11.04} & \underline{11.32} & \underline{1.37} & \underline{13.11} & \underline{0.86} & 1.95 \\
        \cmidrule(lr){2-8}
        & Ours (fine-tuned)         & \textbf{12.11} & \textbf{12.42} & \textbf{1.24} & \textbf{12.17} & \textbf{0.89} & \underline{2.20} \\
        \midrule
        \multirow{6}{*}{1 kHz}
        & Uniform-Sinc              & 0.69 & 3.62 & 7.20 & 61.17 & 0.58 & 1.26 \\
        & AudioUNet                 & 3.73 & 5.71 & 3.93 & 33.48 & 0.58 & 1.45 \\
        & NU-Wave 2                 & 4.04 & 5.98 & 2.06 & 16.82 & 0.56 & 1.45 \\
        & AudioSR                   & 3.96 & 5.82 & 6.15 & 52.12 & 0.59 & \textbf{1.78} \\
        & Ours (without fine-tuning)& \underline{5.52} & \underline{6.38} & \underline{1.78} & \underline{15.12} & \underline{0.72} & 1.32 \\
        \cmidrule(lr){2-8}
        & Ours (fine-tuned)         & \textbf{8.52} & \textbf{9.13} & \textbf{1.42} & \textbf{12.90} & \textbf{0.82} & \underline{1.72} \\
        \bottomrule
    \end{tabular}
\end{table*}

The results of ``Ours (without fine-tuning)'' show that the proposed observation front end transfers effectively from LibriSpeech to AISHELL-1. Without any AISHELL-1 fine-tuning, the proposed method achieves the best SI-SDR, SNR, MR-STFT, LSD, and STOI at all three observation rates among the compared systems. At 4 kHz, it improves SI-SDR from 15.18 dB for NU-Wave 2 to 16.61 dB and reduces LSD from 13.54 for AudioSR to 12.12. At 2 kHz, it obtains 11.04 dB SI-SDR and 11.32 dB SNR, outperforming the strongest baselines by more than 1 dB in both metrics. At 1 kHz, it reaches 5.52 dB SI-SDR and 6.38 dB SNR, while the strongest baseline remains at 4.04 dB and 5.98 dB, respectively. These results show that the learned acoustic observations retain useful reconstruction information under a different language and speaker distribution.

The spectral and intelligibility metrics show the same trend. ``Ours (without fine-tuning)'' gives the lowest MR-STFT and LSD among the systems without AISHELL-1 fine-tuning at all observation rates, showing that the formed observations preserve spectral envelopes and band-energy distributions on AISHELL-1. It also gives the highest STOI at all observation rates, with a particularly clear advantage at 1 kHz, where STOI increases from 0.59 for the strongest baseline to 0.72. PESQ shows a different trend. AudioUNet or AudioSR obtains the highest PESQ among the systems without AISHELL-1 fine-tuning, while the proposed method is stronger in waveform fidelity, spectral consistency, and STOI. This difference indicates that pretrained restoration priors can improve certain perceptual scores, whereas the proposed sampling front end provides more faithful low-rate observations for reconstructing waveform and spectral structure.

Fine-tuning on AISHELL-1 further improves the proposed model. After fine-tuning, the proposed method achieves the best result in most metrics across all observation rates. At 4 kHz, SI-SDR increases from 16.61 dB to 17.72 dB, LSD decreases from 12.12 to 11.32, and PESQ increases from 2.47 to 2.73. At 2 kHz, fine-tuning improves all six metrics and raises SI-SDR from 11.04 dB to 12.11 dB. The improvement is most pronounced at 1 kHz, where SI-SDR increases from 5.52 dB to 8.52 dB, SNR increases from 6.38 dB to 9.13 dB, and STOI increases from 0.72 to 0.82. The larger gain at the lowest observation rate indicates that refining the acoustic response space and its readout is especially useful when the available observations are highly limited.

Without fine-tuning, the proposed front end already forms useful observations for AISHELL-1, giving strong reconstruction performance in waveform, spectral, and intelligibility metrics. Fine-tuning further improves the results by aligning the learned acoustic responses and reconstruction mapping with Mandarin speech characteristics. Overall, the experiments show that the proposed observation front end can transfer across speech corpora and can be further improved when target-domain speech is available.

\section{Conclusion}
\label{sec:conclusion}

This paper presented semantic sampling via learnable observation front ends. The proposed front end forms finite-dimensional observations by mapping the waveform into learned acoustic response channels, combining these responses through a constrained semantic observation matrix, and producing low-rate readouts from the resulting observation channels. Experiments on low-rate speech reconstruction show that these observations provide more informative inputs to the reconstruction network than fixed low-rate waveform samples or predetermined low-rate waveforms used by neural restoration systems. The results demonstrate that, under a fixed observation budget, reconstruction quality depends strongly on how the observations are formed before recovery.

This work evaluates semantic sampling through waveform reconstruction, where reconstruction quality directly reflects the information retained in the observations. Future work may extend this framework to other acoustic objectives, such as intelligibility preservation, speaker-related reconstruction, linguistic content recovery, and downstream speech understanding. These directions would allow the observation front end to be trained with objectives beyond waveform fidelity, while retaining the central idea of forming low-rate observations before reconstruction or inference.

\ifCLASSOPTIONcaptionsoff
  \newpage
\fi

\bibliographystyle{IEEEtran}
\bibliography{references}

@article{nyquist1928certain,
  title={Certain topics in telegraph transmission theory},
  author={Nyquist, Harry},
  journal={Transactions of the American Institute of Electrical Engineers},
  volume={47},
  number={2},
  pages={617--644},
  year={1928},
  publisher={IEEE}
}

@article{shannon1949communication,
  title={Communication in the presence of noise},
  author={Shannon, Claude E},
  journal={Proceedings of the IRE},
  volume={37},
  number={1},
  pages={10--21},
  year={1949},
  publisher={IEEE}
}

@article{landau1967necessary,
  author  = {Landau, H. J.},
  title   = {Necessary Density Conditions for Sampling and Interpolation of Certain Entire Functions},
  journal = {Acta Mathematica},
  volume  = {117},
  pages   = {37--52},
  year    = {1967}
}

@article{papoulis1977generalized,
  title={Generalized sampling expansion},
  author={Papoulis, Athanasios},
  journal={IEEE Transactions on Circuits and Systems},
  volume={24},
  number={11},
  pages={652--654},
  year={1977},
  publisher={IEEE}
}

@ARTICLE{unser2002sampling,
  author={Unser, M.},
  journal={Proceedings of the IEEE}, 
  title={Sampling-50 years after Shannon}, 
  year={2000},
  volume={88},
  number={4},
  pages={569-587},
  doi={10.1109/5.843002}}

@article{vetterli2002sampling,
  title={Sampling signals with finite rate of innovation},
  author={Vetterli, Martin and Marziliano, Pina and Blu, Thierry},
  journal={IEEE Transactions on Signal Processing},
  volume={50},
  number={6},
  pages={1417--1428},
  year={2002},
  publisher={IEEE}
}

@article{candes2006robust,
  title={Robust uncertainty principles: Exact signal reconstruction from highly incomplete frequency information},
  author={Cand{\`e}s, Emmanuel J and Romberg, Justin and Tao, Terence},
  journal={IEEE Transactions on Information Theory},
  volume={52},
  number={2},
  pages={489--509},
  year={2006},
  publisher={IEEE}
}

@article{candes2006near,
  title={Near-optimal signal recovery from random projections: Universal encoding strategies?},
  author={Candes, Emmanuel J and Tao, Terence},
  journal={IEEE Transactions on Information Theory},
  volume={52},
  number={12},
  pages={5406--5425},
  year={2006},
  publisher={IEEE}
}

@article{donoho2006compressed,
  title={Compressed sensing},
  author={Donoho, David L},
  journal={IEEE Transactions on Information Theory},
  volume={52},
  number={4},
  pages={1289--1306},
  year={2006},
  publisher={IEEE}
}

@article{candes2006stable,
  title={Stable signal recovery from incomplete and inaccurate measurements},
  author={Candes, Emmanuel J and Romberg, Justin K and Tao, Terence},
  journal={Communications on Pure and Applied Mathematics: A Journal Issued by the Courant Institute of Mathematical Sciences},
  volume={59},
  number={8},
  pages={1207--1223},
  year={2006},
  publisher={Wiley Online Library}
}

@article{candes2008restricted,
  title={The restricted isometry property and its implications for compressed sensing},
  author={Candes, Emmanuel J},
  journal={Comptes rendus mathematique},
  volume={346},
  number={9-10},
  pages={589--592},
  year={2008},
  publisher={Elsevier}
}

@article{baraniuk2010model,
  title={Model-based compressive sensing},
  author={Baraniuk, Richard G and Cevher, Volkan and Duarte, Marco F and Hegde, Chinmay},
  journal={IEEE Transactions on Information Theory},
  volume={56},
  number={4},
  pages={1982--2001},
  year={2010},
  publisher={IEEE}
}

@article{mishali2010theory,
  title={From theory to practice: Sub-Nyquist sampling of sparse wideband analog signals},
  author={Mishali, Moshe and Eldar, Yonina C},
  journal={IEEE Journal of Selected Topics in Signal Processing},
  volume={4},
  number={2},
  pages={375--391},
  year={2010},
  publisher={IEEE}
}

@article{xie2021deep,
  title={Deep learning enabled semantic communication systems},
  author={Xie, Huiqiang and Qin, Zhijin and Li, Geoffrey Ye and Juang, Biing-Hwang},
  journal={IEEE Transactions on Signal Processing},
  volume={69},
  pages={2663--2675},
  year={2021},
  publisher={IEEE}
}

@article{qin2021semantic,
  title={Semantic communications: Principles and challenges},
  author={Qin, Zhijin and Tao, Xiaoming and Lu, Jianhua and Tong, Wen and Li, Geoffrey Ye},
  journal={arXiv preprint arXiv:2201.01389},
  year={2022}
}

@article{gunduz2022beyond,
  title={Beyond transmitting bits: Context, semantics, and task-oriented communications},
  author={G{\"u}nd{\"u}z, Deniz and Qin, Zhijin and Aguerri, Inaki Estella and Dhillon, Harpreet S and Yang, Zhaohui and Yener, Aylin and Wong, Kai Kit and Chae, Chan-Byoung},
  journal={IEEE Journal on Selected Areas in Communications},
  volume={41},
  number={1},
  pages={5--41},
  year={2023},
  publisher={IEEE}
}

@article{shao2024theory,
  title={A theory of semantic communication},
  author={Shao, Yulin and Cao, Qi and G{\"u}nd{\"u}z, Deniz},
  journal={IEEE Transactions on Mobile Computing},
  volume={23},
  number={12},
  pages={12211--12228},
  year={2024},
  publisher={IEEE}
}

@article{ma2023task,
  title={Task-oriented explainable semantic communications},
  author={Ma, Shuai and Qiao, Weining and Wu, Youlong and Li, Hang and Shi, Guangming and Gao, Dahua and Shi, Yuanming and Li, Shiyin and Al-Dhahir, Naofal},
  journal={IEEE Transactions on Wireless Communications},
  volume={22},
  number={12},
  pages={9248--9262},
  year={2023},
  publisher={IEEE}
}

@inproceedings{chai2023rate,
  title={Rate-distortion-perception theory for semantic communication},
  author={Chai, Jingxuan and Xiao, Yong and Shi, Guangming and Saad, Walid},
  booktitle={2023 IEEE 31st International Conference on Network Protocols (ICNP)},
  pages={1--6},
  year={2023},
  organization={IEEE}
}

@inproceedings{kutay2024classification,
  title={Classification-oriented semantic wireless communications},
  author={Kutay, Emrecan and Yener, Aylin},
  booktitle={ICASSP 2024-2024 IEEE International Conference on Acoustics, Speech and Signal Processing (ICASSP)},
  pages={9096--9100},
  year={2024},
  organization={IEEE}
}

@article{fu2025generative,
  title={Generative ai driven task-oriented adaptive semantic communications},
  author={Fu, Yuzhou and Cheng, Wenchi and Wang, Jingqing and Yin, Liuguo and Zhang, Wei},
  journal={IEEE Transactions on Wireless Communications},
  volume={25},
  pages={9078--9093},
  year={2025},
  publisher={IEEE}
}

@article{kingma2013auto,
  title={Auto-encoding variational bayes},
  author={Kingma, Diederik P and Welling, Max},
  journal={arXiv preprint arXiv:1312.6114},
  year={2013}
}

@article{van2017neural,
  title={Neural discrete representation learning},
  author={Van Den Oord, Aaron and Vinyals, Oriol and others},
  journal={Advances in neural information processing systems},
  volume={30},
  year={2017}
}

@article{oord2018representation,
  title={Representation learning with contrastive predictive coding},
  author={Oord, Aaron van den and Li, Yazhe and Vinyals, Oriol},
  journal={arXiv preprint arXiv:1807.03748},
  year={2018}
}

@article{kuleshov2017audio,
  title={Audio super resolution using neural networks},
  author={Kuleshov, Volodymyr and Enam, S Zayd and Ermon, Stefano},
  journal={arXiv preprint arXiv:1708.00853},
  year={2017}
}

@inproceedings{yamamoto2020parallel,
  title={Parallel WaveGAN: A fast waveform generation model based on generative adversarial networks with multi-resolution spectrogram},
  author={Yamamoto, Ryuichi and Song, Eunwoo and Kim, Jae-Min},
  booktitle={ICASSP 2020-2020 IEEE International Conference on Acoustics, Speech and Signal Processing (ICASSP)},
  pages={6199--6203},
  year={2020},
  organization={IEEE}
}

@article{kim2019bandwidth,
  title={Bandwidth extension on raw audio via generative adversarial networks},
  author={Kim, Sung and Sathe, Visvesh},
  journal={arXiv preprint arXiv:1903.09027},
  year={2019}
}

@article{kong2020hifi,
  title={Hifi-gan: Generative adversarial networks for efficient and high fidelity speech synthesis},
  author={Kong, Jungil and Kim, Jaehyeon and Bae, Jaekyoung},
  journal={Advances in neural information processing systems},
  volume={33},
  pages={17022--17033},
  year={2020}
}

@article{kong2020diffwave,
  title={Diffwave: A versatile diffusion model for audio synthesis},
  author={Kong, Zhifeng and Ping, Wei and Huang, Jiaji and Zhao, Kexin and Catanzaro, Bryan},
  journal={arXiv preprint arXiv:2009.09761},
  year={2020}
}

@article{lee2021nu,
  title={Nu-wave: A diffusion probabilistic model for neural audio upsampling},
  author={Lee, Junhyeok and Han, Seungu},
  journal={arXiv preprint arXiv:2104.02321},
  year={2021}
}

@article{liu2021voicefixer,
  title={VoiceFixer: Toward general speech restoration with neural vocoder},
  author={Liu, Haohe and Kong, Qiuqiang and Tian, Qiao and Zhao, Yan and Wang, DeLiang and Huang, Chuanzeng and Wang, Yuxuan},
  journal={arXiv preprint arXiv:2109.13731},
  year={2021}
}

@article{han2022nu,
  title={NU-Wave 2: A general neural audio upsampling model for various sampling rates},
  author={Han, Seungu and Lee, Junhyeok},
  journal={arXiv preprint arXiv:2206.08545},
  year={2022}
}

@article{defossez2022high,
  title={High fidelity neural audio compression},
  author={D{\'e}fossez, Alexandre and Copet, Jade and Synnaeve, Gabriel and Adi, Yossi},
  journal={arXiv preprint arXiv:2210.13438},
  year={2022}
}

@inproceedings{liu2024audiosr,
  title={AudioSR: Versatile audio super-resolution at scale},
  author={Liu, Haohe and Chen, Ke and Tian, Qiao and Wang, Wenwu and Plumbley, Mark D},
  booktitle={ICASSP 2024-2024 IEEE International Conference on Acoustics, Speech and Signal Processing (ICASSP)},
  pages={1076--1080},
  year={2024},
  organization={IEEE}
}

@article{lu2024towards,
  title={Towards high-quality and efficient speech bandwidth extension with parallel amplitude and phase prediction},
  author={Lu, Ye-Xin and Ai, Yang and Du, Hui-Peng and Ling, Zhen-Hua},
  journal={IEEE Transactions on Audio, Speech and Language Processing},
  volume={33},
  pages={236--250},
  year={2025},
  publisher={IEEE}
}

@inproceedings{kulkarni2016reconnet,
  title={Reconnet: Non-iterative reconstruction of images from compressively sensed measurements},
  author={Kulkarni, Kuldeep and Lohit, Suhas and Turaga, Pavan and Kerviche, Ronan and Ashok, Amit},
  booktitle={Proceedings of the IEEE Conference on Computer Vision and Pattern Recognition},
  pages={449--458},
  year={2016}
}

@inproceedings{wang2017trainable,
  title={Trainable frontend for robust and far-field keyword spotting},
  author={Wang, Yuxuan and Getreuer, Pascal and Hughes, Thad and Lyon, Richard F and Saurous, Rif A},
  booktitle={2017 IEEE International Conference on Acoustics, Speech and Signal Processing (ICASSP)},
  pages={5670--5674},
  year={2017},
  organization={IEEE}
}

@article{mousavi2017deepcodec,
  title={Deepcodec: Adaptive sensing and recovery via deep convolutional neural networks},
  author={Mousavi, Ali and Dasarathy, Gautam and Baraniuk, Richard G},
  journal={arXiv preprint arXiv:1707.03386},
  year={2017}
}

@inproceedings{bora2017compressed,
  title={Compressed sensing using generative models},
  author={Bora, Ashish and Jalal, Ajil and Price, Eric and Dimakis, Alexandros G},
  booktitle={International conference on machine learning},
  pages={537--546},
  year={2017},
  organization={PMLR}
}

@inproceedings{zeghidour2018learning,
  title={Learning filterbanks from raw speech for phone recognition},
  author={Zeghidour, Neil and Usunier, Nicolas and Kokkinos, Iasonas and Schaiz, Thomas and Synnaeve, Gabriel and Dupoux, Emmanuel},
  booktitle={2018 IEEE international conference on acoustics, speech and signal Processing (ICASSP)},
  pages={5509--5513},
  year={2018},
  organization={IEEE}
}

@inproceedings{ravanelli2018speaker,
  title={Speaker recognition from raw waveform with sincnet},
  author={Ravanelli, Mirco and Bengio, Yoshua},
  booktitle={2018 IEEE Spoken Language Technology Workshop (SLT)},
  pages={1021--1028},
  year={2018},
  organization={IEEE}
}

@inproceedings{wu2019learning,
  title={Learning a compressed sensing measurement matrix via gradient unrolling},
  author={Wu, Shanshan and Dimakis, Alex and Sanghavi, Sujay and Yu, Felix and Holtmann-Rice, Daniel and Storcheus, Dmitry and Rostamizadeh, Afshin and Kumar, Sanjiv},
  booktitle={International Conference on Machine Learning},
  pages={6828--6839},
  year={2019},
  organization={PMLR}
}

@article{gozcu2018learning,
  title={Learning-based compressive MRI},
  author={G{\"o}zc{\"u}, Baran and Mahabadi, Rabeeh Karimi and Li, Yen-Huan and Il{\i}cak, Efe and Cukur, Tolga and Scarlett, Jonathan and Cevher, Volkan},
  journal={IEEE Transactions on Medical Imaging},
  volume={37},
  number={6},
  pages={1394--1406},
  year={2018},
  publisher={IEEE}
}

@inproceedings{bahadir2019learning,
  title={Learning-based optimization of the under-sampling pattern in MRI},
  author={Bahadir, Cagla Deniz and Dalca, Adrian V and Sabuncu, Mert R},
  booktitle={international conference on information processing in medical imaging},
  pages={780--792},
  year={2019},
  organization={Springer}
}

@inproceedings{wu2019deep,
  title={Deep compressed sensing},
  author={Wu, Yan and Rosca, Mihaela and Lillicrap, Timothy},
  booktitle={International Conference on Machine Learning},
  pages={6850--6860},
  year={2019},
  organization={PMLR}
}

@article{zeghidour2021leaf,
  title={LEAF: A learnable frontend for audio classification},
  author={Zeghidour, Neil and Teboul, Olivier and Quitry, F{\'e}lix De Chaumont and Tagliasacchi, Marco},
  journal={arXiv preprint arXiv:2101.08596},
  year={2021}
}

@article{zhang2021scalable,
  title={Scalable deep compressive sensing},
  author={Zhang, Zhonghao and Liu, Yipeng and Cao, Xingyu and Wen, Fei and Zhu, Ce},
  journal={arXiv preprint arXiv:2101.08024},
  year={2021}
}

@article{chen2022content,
  title={Content-aware scalable deep compressed sensing},
  author={Chen, Bin and Zhang, Jian},
  journal={IEEE Transactions on Image Processing},
  volume={31},
  pages={5412--5426},
  year={2022},
  publisher={IEEE}
}

@article{hochreiter1997long,
  title={Long short-term memory},
  author={Hochreiter, Sepp and Schmidhuber, J{\"u}rgen},
  journal={Neural computation},
  volume={9},
  number={8},
  pages={1735--1780},
  year={1997},
  publisher={MIT press}
}

@article{schuster1997bidirectional,
  title={Bidirectional recurrent neural networks},
  author={Schuster, Mike and Paliwal, Kuldip K},
  journal={IEEE Transactions on Signal Processing},
  volume={45},
  number={11},
  pages={2673--2681},
  year={1997},
  publisher={Ieee}
}

@article{loshchilov2017decoupled,
  title={Decoupled weight decay regularization},
  author={Loshchilov, Ilya and Hutter, Frank},
  journal={arXiv preprint arXiv:1711.05101},
  year={2017}
}

@inproceedings{panayotov2015librispeech,
  title={Librispeech: an asr corpus based on public domain audio books},
  author={Panayotov, Vassil and Chen, Guoguo and Povey, Daniel and Khudanpur, Sanjeev},
  booktitle={2015 IEEE International Conference on Acoustics, Speech and Signal Processing (ICASSP)},
  pages={5206--5210},
  year={2015},
  organization={IEEE}
}

@inproceedings{le2019sdr,
  title={SDR--half-baked or well done?},
  author={Le Roux, Jonathan and Wisdom, Scott and Erdogan, Hakan and Hershey, John R},
  booktitle={ICASSP 2019-2019 IEEE International Conference on Acoustics, Speech and Signal Processing (ICASSP)},
  pages={626--630},
  year={2019},
  organization={IEEE}
}

@article{taal2011algorithm,
  title={An algorithm for intelligibility prediction of time--frequency weighted noisy speech},
  author={Taal, Cees H and Hendriks, Richard C and Heusdens, Richard and Jensen, Jesper},
  journal={IEEE Transactions on Audio, Speech, and Language Processing},
  volume={19},
  number={7},
  pages={2125--2136},
  year={2011},
  publisher={IEEE}
}

@ARTICLE{gray2003distance,
  author={Gray, A. and Markel, J.},
  journal={IEEE Transactions on Acoustics, Speech, and Signal Processing}, 
  title={Distance measures for speech processing}, 
  year={1976},
  volume={24},
  number={5},
  pages={380-391},
  doi={10.1109/TASSP.1976.1162849}}

@book{quackenbush1988objective, 
    title={Objective Measures of Speech Quality},
    author={Quackenbush, Schuyler R. and Barnwell, Thomas P. and Clements, Mark A.},
    publisher={Prentice Hall},
    address={Englewood Cliffs, NJ},
    year={1988}
}

@article{recommendation2001perceptual,
  title={Perceptual evaluation of speech quality (PESQ): An objective method for end-to-end speech quality assessment of narrow-band telephone networks and speech codecs},
  author={Recommendation, ITU-T},
  journal={Rec. ITU-T P. 862},
  year={2001}
}

@article{loshchilov2016sgdr,
  title={Sgdr: Stochastic gradient descent with warm restarts},
  author={Loshchilov, Ilya and Hutter, Frank},
  journal={arXiv preprint arXiv:1608.03983},
  year={2016}
}

@inproceedings{bu2017aishell,
  title={Aishell-1: An open-source mandarin speech corpus and a speech recognition baseline},
  author={Bu, Hui and Du, Jiayu and Na, Xingyu and Wu, Bengu and Zheng, Hao},
  booktitle={2017 20th conference of the oriental chapter of the international coordinating committee on speech databases and speech I/O systems and assessment (O-COCOSDA)},
  pages={1--5},
  year={2017},
  organization={IEEE}
}

\end{document}